\documentclass{ws-procs9x6}

\newcommand{\Slash}[1]{\ooalign{\hfil/\hfil\crcr$#1$}}
\newcommand{\tr}[1]{ \mathrm{Tr}\left[#1\right] }

\begin{document}

\title{Study of Light Scalar Meson Structure in $D_1$ decay\footnote[7]{Contribution to
KMI Inauguration Conference "Quest for the Origin of Particles and the
Universe" (KMIIN), 24-26 Nov. 2011, KMI, Nagoya University.}}

\author{H. Hoshino$^*$, M. Harada and Y. L. Ma}

\address{Department of Physics, Nagoya University,
Nagoya, 464-8602, Japan\\
$^*$E-mail: hhoshino@th.phys.nagoya-u.ac.jp\\
}


\begin{abstract}
We study the quark structure of the sigma meson through the decay of 
$D_1(2430)$ meson by constructing an
effective Lagrangian for charmed mesons interacting
with light mesons based on the chiral symmetry and heavy quark symmetry.
Within the linear realization of the chiral symmetry, 
we include the P-wave
charmed mesons ($D_1(2430)$, $D_0(2400)$) as the chiral partners 
of ($D^\ast$, $D$), and
the light scalar mesons as the chiral partner of 
the pseudoscalar mesons. 
In
the light meson sector, both the $q\bar{q}$ and $qq\bar{q}\bar{q}$
states are incorporated respecting
their different U(1)$_A$ transformation properties. 
We predict the $D_1 \to D\pi\pi$ decay width 
with two pions in the $I=0,\,l=0$ channel,
which can be tested in the future experiment. 
We find that the width increases with the percentage of the $q\bar{q}$ content
in the sigma meson.
\end{abstract}


\bodymatter

\section{Introduction}
The lightest scalar meson ``sigma'' 
is an interesting object which may give a clue to understand the
fundamental problems of QCD such as the chiral symmetry structure, the
origin of mass and so on. 
The mass spectrum of the light scalar meson nonet including the sigma
meson disfavors the $q\bar{q}$ picture but prefers the
$qq\bar{q}\bar{q}$ interpretation.  
If in the nature there are both 
$q\bar{q}$ and $qq\bar{q}\bar{q}$ states,
they mix to give the physical scalar mesons. 

In this write-up, we summarize a main point of Ref.~\refcite{HHMpre}, where we studied the quark structure of the sigma meson in the heavy-light meson decay, $D_1(2430) \to D \pi \pi$, based on the chiral partner structure between ($D^*,D$) and ($D_1(2430),D^*_0(2400)$).
We first determine the sigma meson mass $m_\sigma$ and the $\sigma$-$\pi$-$\pi$ coupling constant $g_{\sigma \pi \pi}$ by fitting the $S$-wave $\pi$-$\pi$ scattering data below 560 MeV, the maximum energy transferred to the two pions in the $D_1 \to D\pi\pi$ decay.
After a construction of an effective Lagrangian for interactions among light mesons and heavy mesons, we show how the $D_1 \to D\pi\pi$ decay width depends on the quark structure of the sigma meson.

\section{Linear Sigma model with two and four-quark states}

We introduce a linear sigma model for three flavor QCD in low energy region by including $3 \times 3$ chiral nonet fields $M$ and $M'$ representing the $q \bar{q}$ and $qq\bar{q}\bar{q}$ states respectively.
These two chiral nonets have the same transformation property under the chiral SU(3)$_L \times$SU(3)$_R$ symmetry: $M^{(')} \rightarrow {g_L} M^{(')} {g_R}^\dagger$,
where $g_{L,R} \in$ SU(3)$_{L,R}$.
On the other hand, they have the following different U(1)$_A$ transformation properties: 
$ M \rightarrow M e^{+2i\alpha } , \, M' \rightarrow M' e^{-4i\alpha },$
with $\alpha$ as the phase factor of the axial transformation.
We decompose $M^{(')}$ as $M^{(')}=S^{(')}+i\phi^{(')}$, where $S(S')$ is the scalar nonet and $\phi(\phi')$ is the pseudoscalar nonet.
In this study, we adopt the following extended linear sigma 
model~\cite{FJSPRD77034006}\vspace{0.001cm}: 
\begin{equation}
\begin{split}
\mathcal{L}_{\rm light} = & \dfrac{1}{2} \text{Tr} \left( \partial _\mu M \partial ^\mu M^\dagger \right) 
+
\dfrac{1}{2} \text{Tr} \left( \partial _\mu M' \partial ^\mu {M'}^\dagger \right) -V_0 \left(M,M' \right) - V_{\rm SB}
,
\label{eq:Llight}
\end{split}
\end{equation}
where the first two terms are kinetic terms of the $q\bar{q}$ and $qq\bar{q}\bar{q}$ fields, $V_0$ is the SU(3)$_L \times$ SU(3)$_R$ invariant potential and $V_{\text{SB}}$ stands for explicite chiral symmetry breaking terms due to current quark masses.
Here we consider the chiral limit case, that is, $V_{SB}=0$.
We distinguish the $q\bar{q}$ and $qq\bar{q}\bar{q}$ states by their U(1)$_A$ charges.
This U(1)$_A$ symmetry is explicitly broken by anomaly.
As a result, physical mesons are given as mixing states of the
$q\bar{q}$ and $qq\bar{q}\bar{q}$ states through mixing matrices, for
example, iso-singlet scalar mesons are 
$f_{pj} = U_{ja} f_a + U_{jb} f_b + U_{jc} f_c + U_{jd} f_d$, 
where $f_{pj}\,(j=1,\cdots,4)$
are the mass eigenstates with mass ordering $m_1 \leq m_2 \leq m_3
\leq m_4 $ while $f_a=(S_1^1+S_2^2)/\sqrt{2}$, $f_b=S_3^3$,
$f_c=({S'}_1^1+{S'}_2^2)/\sqrt{2}$ and $f_d={S'}_3^3$. 
In the following we call the lightest $f_{p1}$ the sigma ($\sigma$) meson. 
Similarly, for
iso-triplet pseudoscalar meson, we have
$\pi _p = \cos \theta _\pi \pi - \sin \theta _\pi \pi'$,
where $\pi_p$ is the phyisical state while $\pi(\pi')$ is $q\bar{q}\,(qq\bar{q}\bar{q})$ state.
In the present analysis, we identify $\pi_p$ as $\pi(140)$.

The above pseudoscalar mixing angle $\theta_\pi$ relates to the pion decay constant and vaccume expectation values $v_2$ and $v_4$ of the $q\bar{q}$ and $qq\bar{q}\bar{q}$ scalar fields, respectively.
In the chiral limit, we have $ F_\pi \cos \theta_\pi = 2 v_2 $ and $ F_\pi \sin \theta_\pi = - 2 v_4$
, where $F_\pi = 130.41$MeV denotes the decay constant of $\pi(140)$.
Using the chiral symmetry structure of the potential $V$ in 
Eq.~(\ref{eq:Llight}) 
[see Ref.~\refcite{HHMpre} for detail], we obtain the 
following relations: 
\begin{eqnarray}
g_{\pi \pi f_j }
&
= &
\dfrac{\sqrt{2}}{F_\pi}
\left[
\cos \theta_\pi
(U_f)_{ja}
-
\sin \theta_\pi
(U_f)_{jc}
\right]
m^2_{f_j}
,
\label{eq:rela_3_2}
\\
g_{\pi\pi\pi\pi}
&
= &
\dfrac{6}{ F_\pi^2 }
\sum_{j=1}^{4}
\left[
\cos\theta_\pi
(U_f)_{ja}
-
\sin\theta_\pi
(U_f)_{jc}
\right]^2
m_{f_j}^2
.
\label{eq:rela_4}
\end{eqnarray}
Making use of the relations (\ref{eq:rela_3_2}) and (\ref{eq:rela_4}) 
together with the orthonomal conditions, 
$\sum_j (U_{ja})^2 = \sum_j (U_{jc})^2 =1$ and $\sum_j U_{ja} U_{jc} =0$, 
we obtain the following sum rules:
\begin{equation}
\begin{split}
\sum_{j=1}^4
\dfrac{g^2_{\pi\pi f_j}}{m^4_{f_j}}
&
=
\dfrac{2}{F_\pi^2}
,
\qquad
\sum_{j=1}^4
\dfrac{g_{\pi\pi f_j}^2}{m^2_{f_j}}
=
\dfrac{1}{3}
g_{\pi\pi\pi\pi}
,
\end{split}
\end{equation}
where $g_{\pi\pi\pi\pi}$ is the four-pion coupling constant.

\vspace{0.1cm}

\noindent
\begin{minipage}[b]{0.5\linewidth}
\quad
We approximate the $\pi$-$\pi$ scattering amplitude bellow $560$MeV 
as a function of the $\pi$-$\pi$-$ \sigma$ coupling and the sigma
mass.
We fit them to the experimental data of the
$\pi$-$\pi$ scattering amplitude.
The best fitted values are obtained as
$m_\sigma = 606 \pm 9 \, \mbox{MeV}$
and
$g_{\sigma\pi\pi} = 2.16 \pm 0.07 ~ \mbox{GeV}$ with
$\chi^2/{\rm dof} = 3.48/12 =0.29$.
We show the best fitted curves in Fig.~1.
\end{minipage}
\hspace{0.01\linewidth}
\begin{minipage}[b]{0.45\linewidth}
 \includegraphics[width=\linewidth]{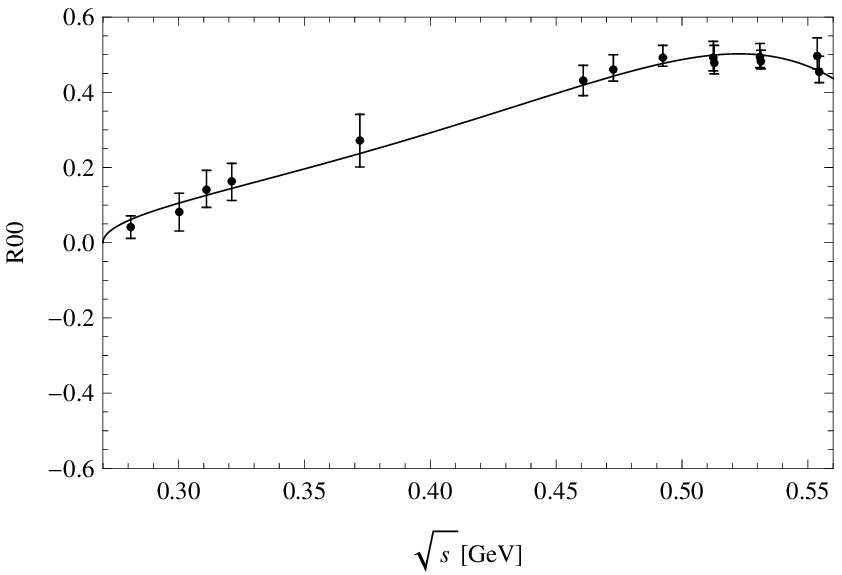}
 {\footnotesize
 Fig.~1.
 { Best fitted curves of the $I=0$, $S$-wave $\pi$-$\pi$ scattering  
 amplitude compared with the experimental 
 data~\cite{Alekseeva:1982uy}. } 
 }
\addtocounter{figure}{1}
\end{minipage}


\section{Effective Lagrangian for the heavy-light mesons}
In this section, we introduce an effective Lagrangian for the heavy mesons coupling to the light mesons based on the heavy quark symmetry combined with the chiral symmetry in order to investigate the mixing effect of the light scalar mesons to $D_1 \rightarrow D\pi\pi$ decay.
In the present analysis, we regard $G=(D_0^*,D_1)$ as the chiral partner of the lowest lying multiplet $H=(D,D^*)$.
We include the $H$ and $G$ doublets into the Lagrangian through
$ \mathcal{H}_R = (G -iH \gamma ^5)/\sqrt{2}, \, \mathcal{H}_L = (G +iH \gamma ^5)/\sqrt{2} $.
The $H$ and $G$ doublets are 
\begin{equation}
\begin{split}
H =  \dfrac{\Slash{v}+1}{2} 
  \left( {D^*_\mu} \gamma ^\mu +iD \gamma ^5 \right) 
,
\quad
G =  \dfrac{\Slash{v}+1}{2} 
  \left( {D_0^*}  -i{D_1}^{\mu } \gamma _\mu \gamma ^5 \right)
,
\end{split}
\end{equation}
with $v^\mu$ being the velocity of the heavy meson. 
The $\mathcal{H}_{L,R}$ fields transform as $\mathcal{H}_{L} \rightarrow \mathcal{H}_L g_L^\dagger$ and $\mathcal{H}_R \rightarrow \mathcal{H}_R g_R^\dagger $.
Then, under the U(1)$_A$ transformation, these fields transform as   
$
\mathcal{H}_L \rightarrow e^{-i\alpha} \mathcal{H}_L 
,
\,
\mathcal{H}_R \rightarrow e^{i\alpha} \mathcal{H}_R 
.
$
We construct the SU(3)$_L \times$SU(3)$_R \times$U(1)$_A$ invariant minimal effective Lagrangian for our study of the $D_1 \to D \pi\pi$ decay as
\begin{equation}
\begin{split}
\mathcal{L}_{\rm heavy} = & \dfrac{1}{2} \tr{ \overline{\mathcal{H}_L} i v \cdot \partial  {\mathcal{H}_L} }
 +\dfrac{1}{2} \tr{ \overline{\mathcal{H}_R} i v \cdot \partial  {\mathcal{H}_R} }
\\
&
-
\dfrac{g_\pi}{4} \tr{M^\dagger \overline{\mathcal{H}_L} {\mathcal{H}_R} + M \overline{\mathcal{H}_R} {\mathcal{H}_L}} 
\\
&
+
i\dfrac{g_A }{2f_\pi} \tr{\gamma ^5 \Slash{\partial }M^\dagger \overline{\mathcal{H}_L} { \mathcal{H}_R} - \gamma ^5 \Slash{\partial } M \overline{\mathcal{H}_R} {\mathcal{H}_L}} 
,
\label{eq:HeavyLag}
\end{split}
\end{equation}
where $g_\pi$ and $g_A$ are parameters.
Using the central values of $\Gamma(D^{*\pm}) = 96 \pm 22$keV and
$\Gamma(D_0^*) = 267 \pm 40$MeV, 
we obtain $\vert g_A \cos \theta_\pi \vert = 0.56$ and 
$ \vert g_\pi \cos \theta_\pi \vert = 3.61$.
Note that we can take $g_A \cos \theta_\pi > 0$ without loss of
generality, but there is a two-way ambiguity for the 
sign of $g_\pi \cos \theta_\pi $.
So, we use $g_A \cos \theta_\pi = 0.56$ and 
$g_\pi \cos \theta_\pi =\pm 3.61$ in the numerical analysis.


\section{Sigma meson structure from $D_1 \to D \pi \pi$ decay}
\begin{figure}[t]
    \begin{minipage}[t]{0.485\linewidth}
		\includegraphics[width=\linewidth]{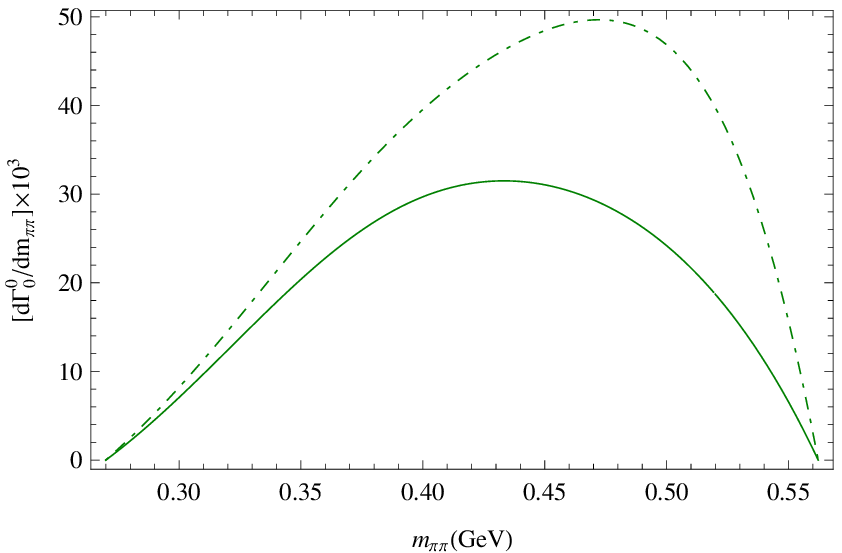}
	\end{minipage}
	\hspace{0.01\linewidth}
    \begin{minipage}[t]{0.485\linewidth}
		\includegraphics[width=\linewidth]{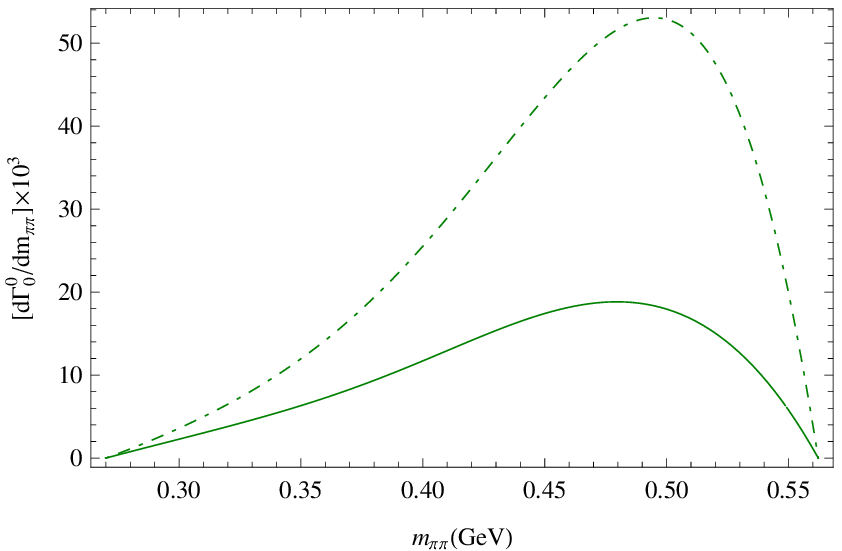}
	\end{minipage}
		\caption{$d\Gamma_{l=0}^{I=0}/dm_{\pi\pi}$ vs $m_{\pi\pi}$ with $h=0$ (solid line), $h=1$ (dashed line). The left panel is for $g_\pi \cos \theta_\pi<0$ and the right one is for $g_\pi \cos \theta_\pi>0$.}		
		\label{fig:2}
\end{figure}

We show the predicted differential decay width $d \Gamma(D_1 \to D \pi \pi)/d m_{\pi\pi}$ with $\pi$-$\pi$ in $I=0$, $l=0$ channel in Fig.~\ref{fig:2} 
for $g_\pi \cos \theta_\pi < 0$ (right
panel) and $g_\pi \cos\theta_\pi >0$ (left panel) with two choices of
the scalar mixing angle $h \equiv U^{-1}_{a1}/\cos \theta_\pi$. 
From this figure one concludes that the differential decay width increases with the percentage of the $q \bar{q}$ component in the sigma meson, and this conclusion dose not change due to the ambiguity of the $g_\pi \cos \theta_\pi$ sign.


\paragraph{Acknowledgments} : 
This work is supported in part by 
the JSPS
Grant-in-Aid 
(GIA) for Scientific
Research on Innovative Areas $\sharp$2104,
Nagoya Univ. GCOE Program QFPU 
from MEXT,
the GIA $\sharp$22224003,
and
National Science Foundation of China $\sharp$10905060.


\end{document}